# Micellar crowding and branching in a versatile catanionic system

Fioretta Asaro[a],[*], Luigi Coppola[b], Luigi Gentile[b],[1]

[a] Department of Pharmaceutical and Chemical Sciences, University of Trieste, via L. Giorgieri 1, Trieste- 34127, Italy
[b] Department of Chemistry, University of Calabria, Arcavacata di Rende (CS)- 87036, Italy

## Abstract

The catanionic system didodecyldimethylammonium bromide (DDAB)-sodium taurodeoxycholate (STDC)-$D_2O$ is characterized by an exceptionally extended $L_1$ region. The comparison of self-diffusion coefficient of the solvent and the DDAB embedded in the micelles provided information about hydra- tion of the aggregates. Moreover, correlating self-diffusion and $^{14}N$ NMR relaxation measurements new insight could be obtained regarding the translational and rotational micellar motions in the crowded solu- tions of systems with 0.2 DDAB-STDC molar ratio. $^{1}H$ 2D NMR spectra gave some hints about the mutual arrangement of DDAB and STDC within the aggregates. For samples with 1.8 and 2.6 DDAB/STD molar ratios the $^{14}N$ data were in agreement with the presence of somewhat branched, interconnected micelles. $^{23}Na$ and $^{81}Br$ dynamic parameters resulted particularly sensitive to the surrounding environment.

The peculiar rheological behaviour of the, highly concentrated, branched micelles samples, namely the steady oscillations in the step tests, was found to be an example of instability originated by the combined effect of material elasticity and slippage at the fixed wall.

## 1. Introduction

The association of the two surfactants didodecyldimethylammoniun bromide (DDAB) and sodium taurodeoxycholate (STDC),

* Corresponding author.
*E-mail addresses:* fasaro@units.it (F. Asaro), luigi.coppola@unical.it (L. Coppola), luigi.gentile@fkem1.lu.se (L. Gentile).
[1] Present address: Division of Physical Chemistry, Lund University, P.O. Box 124, SE-221 00 Lund, Sweden.

which differ under various respects beside charge, gives rise to a variety of intriguing self-assembly motifs, as revealed by Marques et al. [1–4]. STDC is a bile salt (Fig. 1) and the association of bile salts with double-chained surfactants has extensively been investigated in order to understand the assembly of bile salts with phospholipids.

The sodium cholate-lecithin-water system displays a large isotropic phase, $L_1$, region [5]. It becomes by far more extended in the case of DDAB-STDC-$D_2O$ [1]. Noteworthily, detailed rheo- logical and NMR diffusivity measurements enlightened a dramatic microstructural change occurring at the equimolar line, from dis-crete to branched, interconnected micelles [4]. The main subject of the present work are both systems, at high concentrations, at which studies on isotropic systems are not very common. The former kind of aggregates, present at prevailing STDC, are a kind DDAB doped bile salt self-assembly motif and, on the macro scale they are characterized by surprisingly small elastic contributes and viscosities [4]. They allow the study of crowding in a micellar system. At present, translational and rotational diffusion in concentrated systems of colloidal particles has been arising great interest due to its relevance to technological applications and to biology. The high cytoplasm concentration, 30%, which incidentally matches the maximum considered concentration for the micellar system of DDAB-STDC (at 1:5 molar ratio), is postulated to play a paramount role in the vivo functioning of proteins.

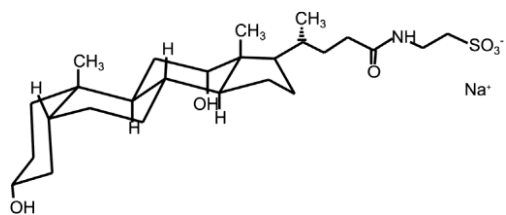

**Fig. 1.** Molecular structure of sodium taurodeoxycholate.

| wt% | surf | mic | $N_{hydr}$ |
|---|---|---|---|
| $W_s = 0.2$ | | | |
| 1 | 0.01 | 0.03 | 42 |
| 5 | 0.04 | 0.12 | 37 |
| 10 | 0.09 | 0.21 | 31 |
| 15 | 0.13 | 0.30 | 28 |
| 25 | 0.23 | 0.48 | 25 |
| 30 | 0.27 | 0.56 | 23 |
| $W_s = 0.4$ | | | |
| 43.3 | 0.41 | | |
| 51.5 | 0.49 | | |
| $W_s = 1$ | | | |
| 65 | 0.63 | 0.93 | 11 |
| 70 | 0.68 | 0.95 | 9 |
| 72.9 | 0.71 | 0.97 | 8 |
| $W_s = 1.8$ | | | |
| 70 | 0.69 | 0.97 | 10 |
| 72.9 | 0.72 | 0.98 | 8 |
| 77.4 | 0.76 | 0.99 | 7 |
| $W_s = 2.6$ | | | |
| 74.5 | 0.74 | 0.99 | 8 |
| 78.5 | 0.78 | 0.99 | 6 |
| 82.2 | 0.81 | 0.99 | 5 |
| 85.3 | 0.84 | 1 | 4 |

Conversely, the systems at high DDAB content, which correspond to branched, somewhat interconnected micelles, are a rather rare self-assembly motif and, on the macro scale they are characterized by surprisingly small elastic contributes and viscosities [4]. They will be investigated in order to get deeper insight into their peculiar rheological behaviour, in particular into the oscillatory step rate response [6].

We make an extensive use of multinuclear NMR spectroscopy because it is an unsurpassed technique to study concentrated, isotropic systems, aiming at gaining information on the systems mainly at the microscopic-molecular level, not much considered in the previous studies. Being dynamics and structure intimately related in complex fluids, the relaxation parameters of the quadrupolar nuclei naturally present in the CTAB systems, namely, $^{14}N$ of DDAB ammonium head-group and $^{23}Na$ and $^{81}Br$ of the monoatomic counterions, will be employed, as well as $^{1}H$ NMR spectra aimed at revealing the mutual arrangement of the two surfactants.

The experimental data are discussed extensively exploiting the analytical tools that are provided by up to date theoretical work. This smart procedure is very convenient to the experimentalist because it allows to gain easily and quickly the significance of the outcomes of the experiments.

## 2. Material and methods

### 2.1. Sample preparation

Sodium taurodeoxycholate (95% purity) and didodecyldimethylammonium bromide (98% purity) were used without any further purifications, deuterium oxide ($^{2}H_2O$) is 99% D. All chemicals were purchased from Sigma-Aldrich.

The composition of the samples, expressed by means of the total surfactant concentration, $C_s$, i.e. (DDAB + STDC) wt%, and of the DDAB/STDC molar ratio ($W_s$) is reported in Table 1 and Fig. S1. The samples were chosen along five dilution lines, i.e. five different $W_s$ were considered.

The mixtures were carefully heated in water bath at 40 °C, and then stirred using a small magnetic stirrer (for the concentrated ones). They were let to stand at room temperature to attain equilibrium for about one month before any measurement. The phase behaviour of the mixtures was checked under cross polarizers and no birefringence was detected. For the NMR measurements, the samples were transferred into 5 mm Pyrex NMR tubes, which were flame-sealed and left for two weeks at 25 °C to equilibrate.

### 2.2. Multinuclear NMR measurements

The $^{1}H$, $^{14}N$ and $^{23}Na$ NMR measurements were carried out at 30 °C on a JEOL Eclipse 400 (9.4 T) NMR spectrometer operating at 399.78 MHz for proton, 107.97 MHz for $^{81}Br$, 105.75 MHz for $^{23}Na$ and 28.89 MHz for $^{14}N$. To record the $^{14}N$ data with a 10 mm probe, the 5 mm tube of the sample was inserted into a 10 mm NMR tube containing water. Temperature was controlled within 0.5 °C by means of a Jeol NM-EVTS3 variable-temperature unit.

Typically spectral widths of 2.8 kHz over 16 K complex points for $^{1}H$, 21.6 kHz over 4 K for $^{81}Br$, 1 kHz over 1 K for $^{23}Na$, and 1 kHz over 256 complex points for $^{14}N$ were used, accumulating 16 scans for $^{1}H$, 16,000 for $^{81}Br$ and 8 for $^{23}Na$, respectively. For $^{14}N$ a variable number of scans, between 500 and 4000, was required to obtain satisfactory spectra depending on the DDAB concentration. For the $C_s = 1\%$ sample, due to the low S/N ratio, 8000 scans were accumulated and the number of data points was halved. Prior to Fourier transformation all the data were zero-filled four times, except the $^{14}N$ ones, which were zero-filled eight times.

The $T_1$ for $^{14}N$ and for $^{23}Na$ were determined by a nonlinear least-squares fit of the inversion-recovery curves obtained using 20 different $\tau$ values and the $T_2$ for $^{23}Na$ were measured by means of the Hahn echo experiment employing 20 different $\tau$ value.

The line-width at half-height, $\Delta v_{1/2}$, were provided by the best fit of the line-shape to a Lorenztian function through the Jeol software Delta [7] and the $^{14}N$ $T_2$ values by means of the relation $\Delta v_{1/2} = \pi T^{-1} T_2^{-1}$.

### 2.3. Rheological measurements

Step rate tests were carried out at 25 ± 0.1 °C on a strain-controlled rheometer RFS III (Rheometrics, USA), equipped with a Couette cell (inner radius 17 mm, gap 1.06 mm) and a Peltier system for temperature control. The samples, which had been homogenized and then kept in a thermostated bath at 25 °C for 24 h, were equilibrated for at least 20 min after filling the geometry and prior to conducting any measurement.

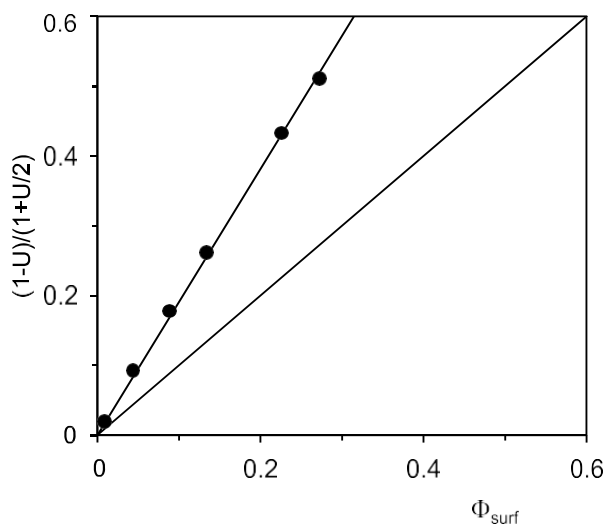

**Fig. 2.** Analysis of water diffusion according to ECM. Plot of $(1-U)/(1+U/2)$, with $U$ defined by Eq. (1), (●) along with the linear fit and the straight line with unity slope (diagonal of the square).

## 3. Results and discussion

### 3.1. NMR diffusometry

First, the translational diffusion coefficients both of water and of DDAB [4] were examined in depth to gain information about self-assemblies hydration and micellar shape.

The data of the $W_s = 0.2$ samples are discussed in terms of discrete micellar aggregates, on the basis of previous studies [4].

In micellar systems, water diffusion is hindered by obstruction, due to the presence of aggregates, as well as by hydration. At low-moderate surfactant concentration, the effective cell model (ECM) describes successfully both effects, [8], and provides a simple test for the micellar shape, its eventual evolution and the relevance of micellar hydration [9]. The common procedure consists in plotting a function of the measured $D_w$ values versus the volume fraction of obstructing objects [9,10]. The function is $(1-U)/(1+U/2)$, where $U$ has the following analytical expression:

$$U = \frac{D_w}{D_{w0}}\left(1 - \frac{\tilde{\phi}_{surf}}{}\right) \quad (1)$$

with $D_{w0}$ the self-diffusion coefficient of free water, here approximated by the one of pure water, and $\tilde{\phi}_{surf}$ the surfactant volume fraction. We estimated it by the densities reported in literature, namely, 1.3 for STDC [11,12] and 1.1 for DDAB [13]. In Fig. 2 the straight line with unity slope, pertinent to the simplest system of obstructing hard spheres, is traced for reference purpose [9].

The data points fit nicely to a straight line with null intercept (Fig. 2), in line with obstruction provided by either hard spheres or weakly prolate micelles, like pure STDC micelles, while marked deviations from linearity occur in the case of oblate micelles with axial ratios exceeding 4–5 [9]. Furthermore, this trend indicates that the micellar shape does not vary across the samples. The slope fairly higher than unity can be rationalized by a strong hydration of micellized surfactant molecules [9,10].

The water self-diffusion coefficients values, $D_w$, obtained from the exponential fit of the echo decay of the D$_2$O residual protons, actually, are average values. This is due to proton exchange, fast on the time-scale of the NMR diffusion experiment, among various sites: namely free and hydrating water and STDC OH groups. In the field of surfactant self-assemblies it is customary working in the convenient two site view [14], i.e. differentiating only between free and "bound" exchangeable protons, with $X_{free}$ and $X_{bound}$ the relevant molar fractions. $D_w$ is accordingly written as follow:

$$D_w = (1 - X_{bound})D_{wfree} + X_{bound}D_{wbound} \quad (2)$$

Eq. (2) enables to estimate the amount of bound water in each sample. We approximated $D_{wbound}$ by the self-diffusion coefficients measured for the surfactant, $D_s$, under the assumption the surfactant molecules are exclusively residing in the micelles, and we corrected the self-diffusion coefficient of "free water" for the presence of micelles in terms of the obstruction of spheres in cubic cell arrangement, through: $D_{wfree} = D_{w0}/(1 + \tilde{\phi}_{surf}/2)$. Exclusively $\tilde{\phi}_{surf}$ has to be considered as the obstructing volume because of the fast exchange between "bound" and "free" water. Actually, the concept of "bound" water is simplistic and misleading, albeit very common because convenient to the experimentalist to quantitatively account for rotational and translational diffusion coefficients exceeding those predicted from the dry volumes of colloidal particles. Instead, this occurrence must not be attributed to true firmly bound water molecules, but rather to the sheath of water molecules, surrounding the particles, with properties quite different from bulk water and with which it is in fast exchange [15]. Nevertheless, in the following we will refer to this "perturbed water" as hydration water, according to the common usage. The number of water molecules per surfactant molecule, usually termed hydration number, $N_{hydr}$, obtained by the above procedure is reported in Table 1. It decreases from 42 to 23, with increasing surfactant concentration, suggesting a tightening of the micelles with concentration. These values are perfectly in line with literature data. Indeed, the peculiar self-assembly in bile salts micelles implies high hydration. The structural determinations and sedimentation experiments afforded a $N_{hydr} \sim 50$ [16,17]. 50 has to be considered as a maximum value since $N_{hydr}$ decreases with concentration [16,18]. The minimum $N_{hydr}$ value reported is 20, for sodium cholate micelles [19]. The $N_{hydr}$ literature values of conventional surfactants, such as DDAB, also depend on the employed method and the micellar model. In dielectric relaxation spectra the water molecules with reduced dynamics, that is with properties different from bulk, include molecules also beyond the closest polar shell. Interestingly, the micellar volumes including them are in agreement with hydrodynamic volumes [20]. For dodecyltrimethylammonium bromide, 10 slow reorienting in addition to 11 frozen water molecules (summing up to 21 perturbed water molecules) were found in highly diluite solutions, and, again, their number decreases with concentration [20]. Eventually, the $N_{hydr}$ of the DDAB/STDC mixed micelles compares well with the $N_{hydr}$, ranging between 39 and 18, of bile salt/phospholipid aggregates [21].

We determined the hydration extent of the samples with $W_s$ 1 in an analogous way. However, since the high concentration of the samples, exceeding 50%, rules out the simple formula used above, which relies on the first order truncation of the series expansion of obstructing volume [14], we employed the analytical expression of Rayleigh for the obstruction provided by a disordered arrangement of spheres, which holds also at higher volume fractions (Eq. S1) [22]. The $N_{hydr}$, reported in Table 1, are neatly lower than for the $W_s = 0.2$ micelles, in line with a higher abundance of the less hydrated DDAB and, above all, to the scarce hydration of very concentrated surfactants' systems. The result that almost all water is interacting with the surfactants is reasonable in such concentrated systems, where, on the other hand the assumption of free water behaving as bulk water, is quite rough.

The experimental self-diffusion coefficient of the surfactant, $D_s$ [4], actually, corresponds to the one of DDAB, since it was determined from the echo decay of the strongest surfactants' proton signal, originated by most protons of DDAB alkyl chains. For the $W_s = 0.2$ samples, it may be safely considered as the self-diffusion

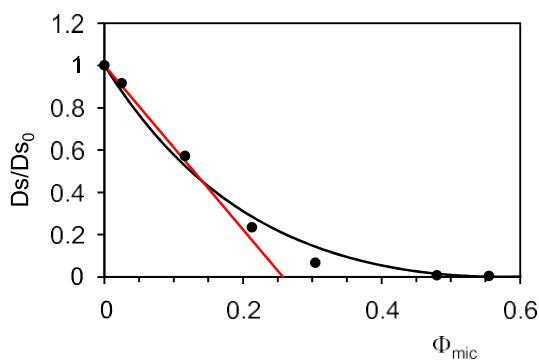

**Fig. 3.** Dependence of micelles' self-diffusion coefficient on hydrated micelles volume fraction: (●) $W_s = 0.2$ experimental datapoints, red straight line: anulus model for $x_1 = 1.5$, black curve: Tokuyama model. (For interpretation of the references to colour in this figure legend, the reader is referred to the web version of this article.)

coefficient of the whole aggregates, owing to the extremely low water solubility of unimeric DDAB.

The diffusivity of micellised surfactants decreases with concentration. We obtained the self-diffusion coefficient of the mixed micelles at infinite dilution, $D_{s0}$, by extrapolating to zero concentration the $D_s$ of the two most diluted samples. The $D_{s0}$ value of $10.33 \cdot 10^{-11}$ m$^2$ s$^{-1}$ points to small micelles, with a hydrodynamic radius, $R_h$, of 1.9 nm, from the Stokes Einstein relation. It is comparable to the one found for STDC micelles, in the presence of a very low concentration of NaCl, which are slightly larger than in pure water [23], and are in line with the previous results for mixed STDC-DDAB micelles [1].

The decrease of $D_s$ with concentration may be attributed both to aggregates interactions and to aggregates' growth. The interactions among charged micelles are not only electrostatic repulsion, but also van der Waals attracting forces, hydrodynamic interactions, mediated by solvent, and self-obstruction. In the present catanionic samples counterions were not removed and they exert a screening effect. When electrostatic repulsion, overwhelming at low ionic strength, is efficaciously screened, the net effect is attraction [23]. In the following we compare the experimental $D_s$ with analytical expression derived on the basis of two models that account for such interactions and/or for anisotropic particle diffusion.

The former is the simplifying description known either as effective hard sphere or as annulus model [24,25]. It is of widespread use and was already validated for bile salt micelles [23,26]. It considers spherical Brownian particles with hydrodynamic interactions and a hard core radius, increased with respect to the hydrodynamic one, to account for the electrostatic repulsion. $x_1$ ($x_1 > 1$) is the ratio of the effective hard core radius and the hydrodynamic radius of the particles. The dependence of $D_s/D_{s0}$ on concentration is linear, with slope function of $x_1$. This linear relationship holds for dilute systems, i.e., $\bar{\nu}_{mic} \cdot x_1^3 \pm 1$. The value of $x_1$ best suited for the STDC-DDAB micelles is 1.5 (Fig. 3), that compares well with 1.2, found for pure STDC micelles by the robust procedure of the combination of the results of two independent techniques, namely both self- and collective diffusion coefficients, obtained by means of PGSE-NMR and quasielastic light scattering, respectively [23].

Then, in order to extend the discussion to higher concentration, we considered the so called Tokuyama model, derived for a concentrated hard sphere suspension of particles with both direct and hydrodynamic interactions [27,28] (Eq. S2). The Tokuyama model had been validated in crowded solutions of proteins by experimental NMR self-diffusion measurements [29] and by sophisticated, atomic detail, Brownian dynamics simulations [27]. These results stress the more important role of hydrodynamic interactions, with respect to direct ones, in determining long term translation diffusion for concentrated solution of interacting particles. Here, we prove that the Tokuyama model is suited also in the case of crowded surfactant self-assemblies.

The $x_1$ slightly higher than for pure STDC micelles, in spite of the expected lower negative charge of micelles due to DDAB incorporation, and the data points sitting slightly below the Tokuyama curve, like in the case of non globular protein [30], may be just symptoms of too crude approximations regarding the micellar volume fraction, and/or the micellar shape, taken as spherical.

### 3.2. NMR measurements

The information about the tumbling of the aggregates may be afforded by the NMR relaxation parameters of $^{14}$N of DDAB, firmly embedded in the aggregates. The $^{14}$N NMR spectra display just the signal of the DDAB head-group, probably being the $^{14}$N resonance of STDC too broad to be detected. At a given micellar tumbling rate, the line-width at half-height, $\Delta\nu_{1/2}$, of the $^{14}$N signals is determined owing to quadrupolar relaxation, by the square of the quadrupole coupling constant, $x$. Since $^{14}$N $x$ of 3.0 [31] and of 0.190 MHz [32], that is 15 times lower, were reported for STDC and DDAB, respectively, it is reasonable to expect $^{14}$N lines more than 200 ($15^2 = 225$) times broader for STDC than DDAB. The $\Delta\nu_{1/2}$ of DDAB signals are collected in Table S1. The transverse relaxation rate values, $R_2$, obtained from the DDAB $^{14}$N $\Delta\nu_{1/2}$ values are plotted in Fig. 4b versus total surfactant concentration, next to the analogous plot of the longitudinal relaxation rates, $R_1$, obtained by inversion recovery experiments. Both $R_1$ and $R_2$ increase with $C_s$ and $W_s$. Steeper slopes in the trends of $R_2$ and $R_1$ with $C_s$ clearly differentiate samples with $W_s > 1$ from those with lower DDAB-SDTC molar ratio. $R_2$ responds much more sensitively than $R_1$ to compositional changes (the span of the y axis of Fig. 4b is 10 times the one of Fig. 4a). $R_2$ is affected by slow motions, which in systems of discrete micelles are mainly surfactant diffusion over the surface of the aggregate and micellar tumbling, therefore $R_2$ afford information about whole self-assemblies. On the contrary, $R_1$ is determined mainly by fast, single molecule reorientational motions (Fig. 4a).

The effect of fast motions can be removed from $R_2$ by subtracting $R_1$. $\Delta R$ ($\Delta R = R_2 - R_1$) provides immediate info about slow motions, since it is directly proportional to the correlation time of slow motion, $T_c^s$ (Eq. S10), through the product of quadrupole coupling constant, $x$, and order parameter, S (see Supporting Information). A crude estimate of $|Sx|$ is provided by the ones of liquid crystalline phases, where it is easily obtained from the residual quadrupolar splitting, $\Delta\nu_Q$, of the NMR signal ($|Sx| = 4/3|\Delta\nu_Q|$). In the following we will make reference to a value of 11 kHz for $^{14}$N $|\Delta\nu_Q|$, since the values reported in literature for dialkyldimethylammonium bromides are 10.3 kHz in an inverse hexagonal phase and 11.9 kHz in a lamellar phase of DDAB [33], and

12.62 and 12.27 kHz, in a lamellar phase of dimethyldidecylammonium bromide at 0.22 mol fraction, at 300 and 305 K, respectively [34].

The linear extrapolation to $C_s = 0$ of the measured $\Delta R$ values for 5 and 10% $C_s$ samples returns $\Delta R_0 = 6.4$ s$^{-1}$, and a $T_c^s = 7$ ns, from it. This value is in line with the expected rotational correlation time for a sphere of 1.9 nm radius, from $D_{s0}$. The comparison of relaxation data and self-diffusion coefficients indicates that micellar tumbling is the main slow motion determining $^{14}$N transverse NMR relaxation of $W_s = 0.2$ samples.

The steeper slope of $^{14}$N $\Delta R$, and consequently of $T_c^s$, with $C_s$ observed for $C_s \geq 20\%$, may be attributed to micellar growth and intermicellar interactions. The longest $T_c^s$, relevant to the most concentrated $W_s = 0.2$ sample ($C_s = 30\%$), exceeds 250 ns. It is on the same order of magnitude of the $T_c$ for the tumbling of the spherocylindrical micelles of decylammonium chloride 20–30% in water [35] and is reflecting the Brownian motion restricted in space owing to the presence of nearby aggregates.

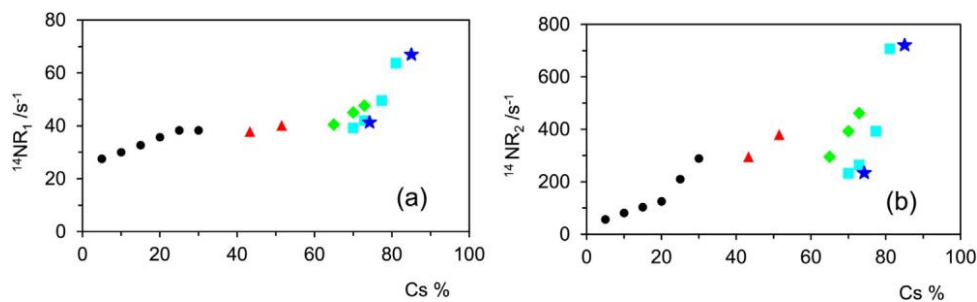

**Fig. 4.** Trends of [14]N longitudinal relaxation rate ($R_1$) (a) and of [14]N transverse relaxation rate ($R_2$) (b) with the total surfactant concentration (C$_s$). W$_s$ = 0.2 (●), 0.4 (▲), 1 (◆), 1.8 (■), 2.6 (★).

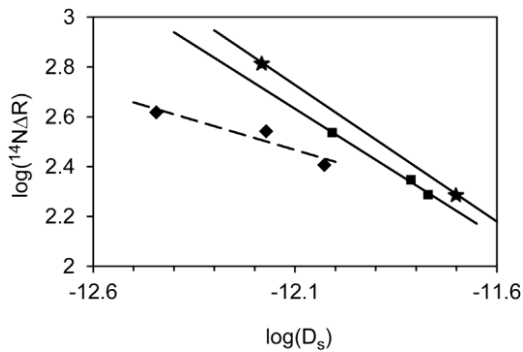

**Fig. 5.** Plot of the decimal logarithm of [14]N ◆R vs. that of $D_s$ for W$_s$ = 1 (+), 1.8 (●) and 2.6 (*) samples and linear fits.

For spherical particles rotational diffusion is less severely affected by caging than translational diffusion measured at times longer than the interval between collisions with nearby particles, which is the case of PGSE NMR experiments. This is clearly reflected, for example, by the different concentration dependences of translational and rotational dynamics of spherical proteins [36]. The same holds for the present micelles, for which $D_s$ is more sensitive to concentration changes [4] than [14]N ◆R, which mainly reflects rotational reorientation. Actually above C$_s$ = 10%, the micelles becomes elongated, so that both translational diffusion and viscosity are strongly enhanced. Their aspect ratio is moderate because the aggregate's tumbling, as reported by [14]N ◆R, is not as strongly hindered and no liquid crystalline order is observed, at least at room temperature, even for the highest concentration, 30%.

$R_1$ weakly increasing with W$_s$ indicates a slight enhancement of DDAB molecular motional freedom.

Along each series $R_2$ increases with C$_s$ and, interestingly, it is lower at corresponding C$_s$ but higher W$_s$. This might be the effect of faster lateral diffusion, as suggested by the analogous DDAB diffusivity trends [4]. A simple test to check the correspondence between surfactant lateral diffusion and the slow motion dominating transverse relaxation is the log–log plot of [14]N ◆R vs. $D_s$ [37]. Indeed a nice linear correlation with minus unity slope can be envisaged for both W$_s$ = 1.8 and 2.6 series in Fig. 5, in which are reported, for comparison purpose, also the data of W$_s$ = 1 samples, which are best fitted to a straight line with lower slope.

The relationship ensues from the inverse dependence of $T_c$s on surface surfactant diffusion, $D_{lat}$ ($T_c$s = 1/(6$D_{lat}$)). In bicontinuous cubic phases $D_{lat}$ and macroscopic surfactant diffusion, $D_s$, are simply related as $D_s = 2/3\ D_{lat}$ for continuous surfaces and $D_s = 1/3 D_{lat}$ for interconnected cylinders [38]. The present observation are in agreement with the microstructure of branched, interconnected micelles previously suggested on the basis of self-diffusion and rheological parameters, which also ruled out a true bicontinuous structure [4]. This is a further example of how NMR relaxation parameters affected by typical slow dynamics of the lyotropic organization, although not able to afford the immediate information of the mesostructure kind, can be exploited in conjunction with independent experimental data to discriminate among different structural arrangements. Unfortunately, in the case of sparsely interconnected micelles, at variance with a fully interconnected network, the relationship between $D_{lat}$ and $D_s$ is not straightforward, due to the different sensitivity to defects [37]. Nevertheless, it can be stated that the fast DDAB self-diffusion [4], measured by PGSE NMR, on macroscopic length and time scales, is associate with fast self-diffusion on molecular length and time scales.

Biexponential decays for the magnetization of [23]Na, which possesses a nuclear spin quantum number of 3/2, are expected in the presence of slow motions. However, in all the investigated systems any deviation from a single exponential curve was detected neither for longitudinal nor for transverse magnetization. This and the closeness of $R_1$ and $R_2$ values (in Fig. 6a and b, the y axes are spanning the same range) can be explained by strong sodium dissociation, enhanced by the presence of the DDAB cation in the aggregates. The observed $R_1$ and $R_2$ values are determined mainly by "free" sodium, under the assumption of a fast exchange of sodium among the various sites. Both relaxation rates increase upon increasing surfactant concentration with progressively steeper slopes (Fig. 6). The faster relaxation rates, especially the $R_2$, for samples with C$_s$ > 60% reflect slower motions but, above all, an enhanced sensitivity to them through a stronger EFG at the [23]Na nucleus. In facts the EFG at monoatomic ions nuclei are due by external charges and dipoles and therefore extremely sensitive to environmental changes, hydration included.

Again ◆R is the appropriate parameter, because in this way the contribution of free sodium is removed, provided it is the same for $R_1$ and $R_2$. Actually, ◆R/0.3 would be even better because more directly related to the low frequency spectral density [39]. At variance with [14]N ◆R, which responds well to C$_s$ also in the W$_s$ = 0.2 and 0.4 series, the [23]Na ◆R (Table S2) increases dramatically in the W$_s$ 1, 1.8 and 2.6 series. Thus [23]Na appears to reflect well the environmental changes in the most concentrated samples with low water content due to the intermolecular origin of its EFG.

Bromide is the didodecyldimethylammonium counterion. The [81]Br signals could be exclusively observed for the W$_s$ = 0.2 samples whereas for the other series they were too broad to be detected.

The counterion Br$^-$ plays a special role in the self-assembling of didodecyldimethylammonium surfactants, due to its ability to interact more strongly than other, less polarizable counterions [40]. The large line-widths observed for [81]Br (Table S2) suggest a significant interaction, in spite of the overall negative charge of the aggregates.

[1]H NMR spectroscopy affords information inside the aggregate, at molecular and submolecular detail.

The TDC$^-$ signals can be better appreciated in the [1]H NMR spectrum of the W$_s$ = 0.2, samples, e.g., lowest trace (Fig. 7), whereas those of DDAB in the uppermost trace, $W_s = 2.6$, where DDAB is by far more abundant. The proton chemical shifts do not vary significantly within each series; while internal changes were observed among different series. In particular the most sensitive are DDAB protons of the groups directly bonded to nitrogen and Me21, the methyl of the side chain of STDC (the detailed assignment of proton signals is reported as Supporting Information). Unfortunately most of the STDC signals cannot be observed at the higher concentrations and $W_s$, due to the extended overlap of the signals of steroidal nucleus and to their progressive, severe broadening (Fig. 7).

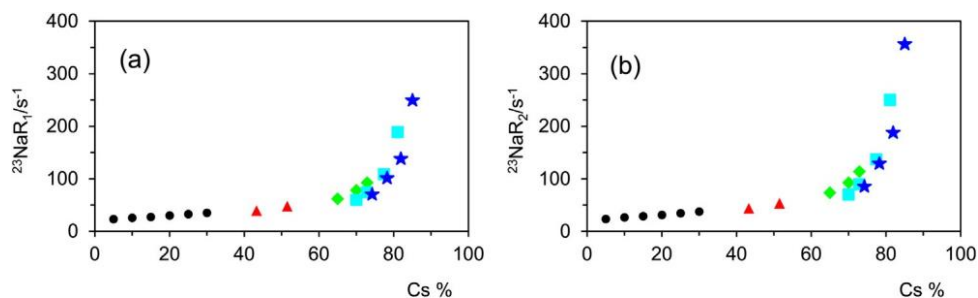

**Fig. 6.** Trends of $^{23}$Na longitudinal relaxation rates ($R_1$) (a) and transverse relaxation rates ($R_2$) (b) with the total surfactant concentration ($C_s$). $W_s = 0.2$ (●), 0.4 ( ▲ 1 ( 🠗 1.8 (■), 2.6 ( ★ ).

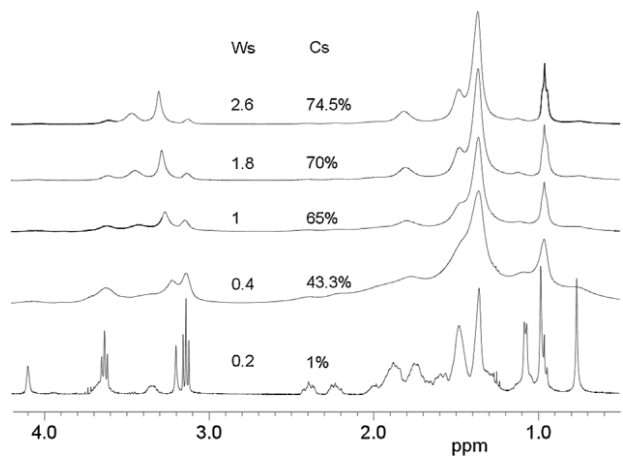

**Fig. 7.** $^1$H NMR spectra of samples with increasing DDAB to STDC molar ratio and total concentration.

Across the $W_s = 0.2$ samples, (Table S3 and Fig. S3) the protons next to taurine $SO_3^-$ head-group show the lowest line-width indicating high motional freedom, able to counteract the broadening effect of crowding, which hampers micellar tumbling. The absence of noticeable intermolecular cross-peaks in the NOESY spectra (Fig. S4) of the sample with $W_s = 0.2$ and $C_s = 5\%$ is in agreement with the remarkable flexibility and motional freedom of the side chain of the bile salts in the helical structure of bile salts micelles [41,42], indicating that the solution structure of pure STDC aggregates is maintained in the series of samples with lowest DDAB content.

In the $W_s > 0.2$ series, the concentration broadening effect on proton signals line-width is even more pronounced, especially for the protons of the STDC steroidal skeleton and of DDAB N–CH$_2$. On the contrary for $W_s \geq 1$, the DDAB alkyl chains proton envelope is rather narrow indicating that the tails are fairly mobile, and therefore in a molten state. The resonance of the methyl groups of the DDAB alkyl chains are not very broad, as well, indicating that also the methyl rotation about the N–C axis is not much hindered. On the contrary, the N–CH$_2$ groups experience less conformational freedom due to the packing requirements for double-tail surfactants residing in an aggregate [34] of the kind of conventional micelles, where the surfactant head-groups are anchored at the aqueous interface. The correlation peak between the protons of STDC taurine and methyls of the DDAB head-group that might be visualized in the NOESY spectrum of sample with $W_s = 1$ and $C_s = 65\%$ (Fig. S6), at variance with the previous $W_s = 0.2$ sample, albeit very weak, is a further hint that a drastic change in the micellar structure has occurred. This hypothesis is supported by the observation, reported in literature, for the catanionic pair cetyltrimethylammonium bromide (CTAB) and sodium deoxycholate (SDC), at 1:1 molar ratio, of a cross-peak in the NOESY and ROESY spectra between the methyl, next to the carboxylate group of deoxycholate, and CTAB ammonium methyls, suggesting that the polar head of CTAB is close to the negatively charged terminal group of bile salt side chain [43].

The lack of further intermolecular NOE signals does not allow to assess the precise localization of STDC inside the aggregates. The proximity of DDAB head-group to STDC side chain and the mobility of DDAB alkyl chains are features in common with lecithin-bile salts mixed micelles. Also in that case the molecular orientation of STCD in the aggregate has to be cleared. The presently, generally accepted description was formulated by Hjelm [44] on the basis of SANS data from a system made of egg yolk dipalmitoylphosphatidylcholine (DPPC) and sodium taurocholate. It consists of cylindrical mixed micelles [45,46], in which the DPPC molecules are arranged radially with the head-group outwards and in close association with the bile salt head-group. The bile salts impart the necessary curvature acting as wedges. Less clear is the orientation of the bile salt molecule: whether radial, that is with the long axis perpendicular to the long axis of the rod [44], or lying on the surface [46], considering that in bile acid/phosphatidylcholine mixed monomolecular layers the bile acid molecules have the long axis of the steroidal nuclei parallel to the air-water interface [47]. The surface localization would better agree with the high curvature of rodlike aggregates [48].

### 3.3. Rheology

The most interesting rheological feature is the steady periodical modulation of the step rate response curve (stress vs. time at constant shear rate in the linear viscoelastic region) observed for the $W_s \geq 1$ samples at 6° C, temperature at which they exhibited a shear thinning behaviour [6].

Here we carried out the transient experiments at 25 °C, where all the samples behave linearly, the viscosity being constant over the whole range of accessible shear rate [4], nevertheless, the phenomenon appeared again. Selected examples are reported in Fig. 8. The oscillations do not occur in the samples with low $W_s$ and $C_s$.

The modulation of the step rate response curves is weak in the case of the $W_s = 1$, and quite strong for the $W_s > 1$ samples (Fig. 8). It is periodic, with a frequency of 0.2 Hz almost constant across the various samples, as proven by means of Fourier transform, and close to that reported at low temperature, in spite of quite different zero shear viscosities [6]. The amplitude, on the contrary, varies sen-sitively. About the possible origin of this phenomenon, it should be considered that flow instabilities have a manifold of causes, the main ones being shear induced reorganization of the fluid microstructure, inertia and elasticity, and interfacial instabilities, as explained in a recent review [49]. Oscillating shear response is well known in the case of shear banding systems that, upon shearing, separate into bands of different viscosity and microstructure and/or concentrations. In such systems, either the relative abundance of the bands may periodically vary in time due to a feedback triggered by stress [50] or one of the bands may host visco-elastic instabilities [51–53]. Among surfactant self-assemblies, wormlike micelles are the best known examples of shear banding fluids [51,54]. Their rheological behaviour is characterized by sizeable induction time for the oscillations, required by the reorganization of the microstructure, and, above all, a non-monotonic flow curve [51]. This is not the case of the present systems at 25 °C and, therefore, the oscillatory response is, rather, likely related to their, albeit weak, elastic properties. An pure elastic instability is not probable because of the low Weissenber number (Wi = $\dot{y}\,T_R$) [55], which is expression of fast structural relaxation. The relaxation time, $T_R$, should be very short, since enough high shear rates to determine it from the flow curve at room temperature were unattainable.

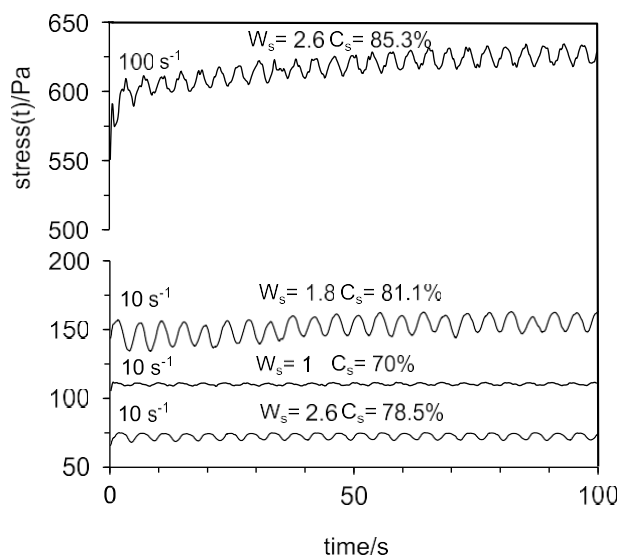

**Fig. 8.** Selected step-rate tests.

On the other hand, theoretical calculations carried out on an Oldroyd-B fluid model with non linear slip at the fixed wall showed that elasticity can induce the oscillatory response to a sudden velocity change of the mobile wall even at very low Wi values. The perturbation is instantaneous and its period is independent of the final velocity of the moving wall [56,57], as happens in our systems. Thus material elasticity and slippage at the fixed wall altogether provide a plausible explanation for the oscillations of the shear step response.

## 4. Conclusions

A detailed analysis of $D_w$ and $D_s$ experimental data led to the determination of the amount of hydration water. Here it must be intended as the amount of water with characteristics different from bulk water that enters in determining the transport properties of micelles. Even if caution must be exerted, in extending it to properties, not straightforwardly correlated to transport, it looks stimulating to investigating the link, if any, between this kind of hydration water and the water with properties, not straightforwardly related to transport phenomena, perturbed with respect to

bulk, e. g. reactivity, which is relevant to catalysis by micelles, a hot topic in green chemistry

The present discrete micelles of the $W_s$ = 0.2 series appear to be a well suited micellar model systems for studying the effect of caging by surrounding aggregates on micellar translational and rotational diffusion, exploiting surfactant self-diffusion coefficients and $^{14}$N NMR relaxation parameters. The Tokuyama model, widely employed in the calculation of self-diffusion coefficient in crowded protein solutions, resulted fairly adequate also for the discrete micellar aggregates in a very wide concentration range.

For the samples of branched micelles at high concentration and large DDAB content, the $^{14}$N NMR relaxation results indicate that the low macroscopic viscosity and elasticity are paralleled by a fast surface diffusion of DDAB at the microscopic level. The notable oscillatory behaviour in the flow curve can be explained by means of material elasticity and slippage at the fixed wall.

## Acknowledgements

This work was supported by Italian MIUR (Project PRIN 2010BNZ3F2_006 "Development of Energy-targeted Self-assembled supramolecular systems: a Convergent Approach through Resonant information Transfer between Experiments and Simulations – DESCARTES").

## Appendix A. Supplementary data

Supplementary data associated with this article can be found, in the online version, at http://dx.doi.org/10.1016/j.colsurfa.2016.06.035.

# Glossary

*DDAB*: Didodecyldimethylammonium bromide
*STDC*: Sodium taurodeoxycholate
*PGSE NMR*: Pulsed gradient spin echo NMR
$W_s$: DDAB to STDC molar ratio
$C_s$: Total surfactant concentration
$D_w$: Water self-diffusion coefficient
$D_g$: DDAB self-diffusion coefficient
$D_{s0}$: Micellar self-diffusion coefficient at infinite dilution
$\Delta v_{1/2}$: Line width at half height
Longitudinal relaxation rate $R_1$:
Transverserelaxation rate $R_2$:
$\Delta R$: $=(R_2 - R_1)$
*ECM*: Effective cell model
  *surf*: Surfactant volume fraction
  *mic*: Hydrated micelles volume fraction
$R_h$: Hydrodynamic radius
$x_1$: Ratio between the hard-core radius and the hydrodynamic radius
$N_{hyd}$: Number of hydration water molecules
*CTAB*: Cetyltrimethylammonium bromide
*SDC*: Sodium deoxycholate
*DPPC*: Dipalmitoylphosphatidylcholine
$T_c^2$: Correlation time of slow motions
$x$: Quadrupole coupling constant
$S$: Order parameter
$\Delta v_Q$: Quadrupolar splitting
$D_{int}$: Surface surfactant diffusion
*EFG*: Electric field gradient
*MLV*: Multilamellar vesicle
*Wi*: Weissenber number
$\dot{y}$: Shear rate.